\documentclass[aps,prl,twocolumn,superscriptaddress,amsmath,amssymb]{revtex4-1}
\usepackage[latin1]{inputenc}
\usepackage{bm}
\usepackage{multirow,amssymb,amsbsy,amsmath}
\usepackage{graphicx}
\usepackage{mathrsfs}
\usepackage{verbatim}
\usepackage{amssymb}
\usepackage{amsfonts}
\usepackage{bbm}
\usepackage{amsmath}
\usepackage{lmodern}
\usepackage{appendix}
\usepackage{color,soul}
\usepackage[colorlinks,citecolor=blue]{hyperref}
\usepackage{ulem}
\usepackage{upgreek}

\newcommand{\R}[1]{\textcolor{red}{#1}}

\makeatletter
\usepackage{pifont}
\makeatother

\usepackage{dcolumn}
\usepackage{epstopdf}
\usepackage{subfigure}

\begin{document}
\renewcommand{\figurename}{{\bf Figure}}


\title{Room temperature coherent manipulation of single-spin qubits in silicon carbide with a high readout contrast}

\author{Qiang Li}
\altaffiliation{These authors contributed equally to the work}
\author{Jun-Feng Wang}
\altaffiliation{These authors contributed equally to the work}
\author{Fei-Fei Yan}
\author{Ji-Yang Zhou}
\author{Han-Feng Wang}
\author{He Liu}
\affiliation{CAS Key Laboratory of Quantum Information, University of Science and Technology of China, Hefei, Anhui 230026, People's Republic of China}
\affiliation{CAS Center for Excellence in Quantum Information and Quantum Physics, University of Science and Technology of China, Hefei, Anhui 230026, People's Republic of China.}

\author{Li-Ping Guo}
\author{Xiong Zhou}
\affiliation{Key Laboratory of Artificial Micro- and Nano-structures of Ministry of Education and School of Physics and Technology, Wuhan University, Wuhan, Hubei 430072, People's Republic of China.}

\author{Adam Gali}
\email{gali.adam@wigner.hu}
\affiliation{Department of Atomic Physics, Budapest University of Technology and Economics, Budafoki $\acute{u}$t. 8, H-1111, Hungary}
\affiliation{Wigner Research centre for Physics, PO. Box 49, H-1525, Hungary}

\author{Zheng-Hao Liu}
\author{Zu-Qing Wang}
\author{Kai Sun}
\author{Guo-Ping Guo}
\author{Jian-Shun Tang}
\affiliation{CAS Key Laboratory of Quantum Information, University of Science and Technology of China, Hefei, Anhui 230026, People's Republic of China}
\affiliation{CAS Center for Excellence in Quantum Information and Quantum Physics, University of Science and Technology of China, Hefei, Anhui 230026, People's Republic of China.}

\author{Hao Li}
\author{Li-Xing You}
\affiliation{State Key Laboratory of Functional Materials for Informatics, Shanghai Institute of Microsystem and Information Technology, Chinese Academy of Sciences(CAS), Shanghai 200050, People's Republic of China}

\author{Jin-Shi Xu}
\email{jsxu@ustc.edu.cn}

\author{Chuan-Feng Li}
\email{cfli@ustc.edu.cn}

\author{Guang-Can Guo}

\affiliation{CAS Key Laboratory of Quantum Information, University of Science and Technology of China, Hefei, Anhui 230026, People's Republic of China}
\affiliation{CAS Center for Excellence in Quantum Information and Quantum Physics, University of Science and Technology of China, Hefei, Anhui 230026, People's Republic of China.}

\date{\today}

\begin{abstract}
{\flushleft{\bf{ABSTRACT}}}\\
Spin defects in silicon carbide (SiC) with mature wafer-scale fabrication and micro/nano-processing technologies have recently drawn considerable attention. Although room temperature single-spin manipulation of colour centres in SiC has been demonstrated, the typically detected contrast is less than 2$\%$, and the photon count rate is also low. Here, we present the coherent manipulation of single divacancy spins in 4H-SiC with a high readout contrast ($-30\%$) and a high photon count rate (150 kilo counts per second) under ambient conditions, which are competitive with the nitrogen-vacancy (NV) centres in diamond. Coupling between a single defect spin and a nearby nuclear spin is also observed. We further provide a theoretical explanation for the high readout contrast by analysing the defect levels and decay paths. Since the high readout contrast is of utmost importance in many applications of quantum technologies, this work might open a new territory for SiC-based quantum devices with many advanced properties of the host material.\\

{\flushleft{\bf{Keywords:}}} silicon carbide, single divacancy defects, spin coherent control, high readout contrast, bright photon emission.
\end{abstract}


\maketitle

\noindent
{\bf INTRODUCTION}

\noindent Colour centres in SiC have recently attracted broad interest as electrically driven, highly bright single-photon sources and defect spins with long coherence time~\cite{SPS:REVIEW:SPIN:Wrachtrup:01, SPS:REVIEW:SPIN:Wrachtrup:02, SiC:Review:2020JPP:Castelletto, SiC:Review:2020APL:Son, SiC:Diode1, SiC:Diode2, 4HSiC:SPS:pin:18ACSPhotonics, 4HSiC:SPS:pin:18APL, 4HSiC:VsiVc:single:01, 4HSiCand3CSiC:VsiVc:single:02, 4HSiC:DPS:2020Science, 4HSiC:Vsi:single:SIL:RT:ODMR:PL:2015}. The most widely studied spin defects in SiC are divacancies (missing a silicon atom and an adjacent carbon atom, $\mathrm{V_{Si}V_C}$)~\cite{SiC:VsiVc:Theory1:Gali:2011, 4HSiC:VsiVc:ensemble:Nature:2011, 4HSiC:VsiVc:Single:EOCC, 4HSiC:DPS:2020Science}, silicon vacancies (missing a silicon atom, $\mathrm{V_{Si}}$)~\cite{4HSiC:Vsi:single:SIL:RT:ODMR:PL:2015, 4HSiC:Vsi:EleStructure, 4HSiC:Vsi:single:neutron:ODMR, 4HSiC:Vsi:V1:single:SIL:PLE, 4HSiC:Vsi:Qudits:RT} and nitrogen-vacancy centres (consisting of a nitrogen impurity substituting a carbon atom and a silicon vacancy adjacent to it, $\mathrm{N_{C}V_{Si}}$)~\cite{SiC:NcVsi:proton:PL:16PRB, SiC:NcVsi:Theory:Gali:17PRB, WJF:4HSiC:NV:PRL2020, 4HSiC:NV:GWB:NanoLett2020}, the spin states of which can be optically polarized and readout. Although hundreds of SiC polytypes exist, many works focus on a specific polytype, namely the 4H polytype (4H-SiC), due to its high crystal quality. On the other hand, the $\mathrm{V_{Si}V_C}$ defects in SiC are near-infrared photoluminescence (PL) emissions and have versatile applications, including quantum information processing~\cite{4HSiC:VsiVc:ensemble:entanglement} and multifunctional sensing, such as magnetic fields~\cite{SiC:VsiVc:ensemble:implantation:C:Awschalom}, electric fields~\cite{4HSiC:Vsi:VsiVc:EOCC:electrometry}, strain~\cite{4HSiC:VsiVc:ensemble:sensor:electrical:strain, 4HSiC:VsiVc:Sensor:LocalStrain:25nm}, and temperature~\cite{WJF:4HSiC:VsiVc:ensemble:sensor:LT, YFF:4HSiC:VsiVc:ensemble:sensor:HT}. Moreover, these defect spins can be flexibly controlled by microwaves~\cite{4HSiC:VsiVc:single:01, 4HSiCand3CSiC:VsiVc:single:02}, electronics~\cite{6HSiC:VsiVc:EODMR, 4HSiC:VsiVc:Single:EOCC}, and acoustics~\cite{4HSiC:VsiVc:Ensemble:Acoustics}, which have garnered great interest.

Depending on the location of the vacancies (hexagonal ($h$) and quasi-cubic ($k$)), four identified types of $\mathrm{V_{Si}V_C}$ defects exist in 4H-SiC, namely, $hh$ (PL1), $kk$ (PL2), $hk$ (PL3) and $kh$ (PL4) defects~\cite{4HSiC:VsiVc:ensemble:Nature:2011, SiC:VsiVc:ensemble:implantation:C:Awschalom}. In addition to the four known types of $\mathrm{V_{Si}V_C}$ defects, there are also the PL5, PL6, and PL7 defects~\cite{4HSiC:VsiVc:ensemble:Nature:2011, SiC:VsiVc:ensemble:implantation:C:Awschalom} that have been recently assigned to divacancy configurations inside stacking faults, which act as local quantum wells in 4H-SiC and make PL5-PL7 colour centres robust against photo-ionisation~\cite{4HSiC:PL567:Stable:Theory:Gali}. Thus, we use $\mathrm{V_{Si}V_C}$ to refer to the PL1-PL7 defects. Although room temperature single-spin manipulation of colour centres in SiC has been previously demonstrated~\cite{4HSiC:Vsi:single:SIL:RT:ODMR:PL:2015,4HSiC:Pillars:17NanoLett}, the typical detected contrast is less than 2\%, and the photon count rate is also low, which limits their applications.

In this work, we prepare arrays of single $\mathrm{V_{Si}V_C}$ defects in 4H-SiC through carbon ion ($\mathrm{C^+}$) implantation and annealing. We then investigate the spin and optical properties of single $\mathrm{V_{Si}V_C}$ defects at room temperature. Surprisingly, for single PL6 defects, the single-photon saturated count rate is up to $150$~kcps (kilo counts per second), which is almost 5 times and 15 times higher than that of single PL1-PL4 divacancies~\cite{4HSiC:VsiVc:single:01, 4HSiCand3CSiC:VsiVc:single:02} and single $\mathrm{V_{Si}}$ in bulk 4H-SiC~\cite{4HSiC:Vsi:single:SIL:RT:ODMR:PL:2015, LQ:4HSiC:Vsi:single:array}, respectively. Moreover, the contrasts of the continuous-wave (CW)-optically detected magnetic resonance (ODMR) spectrum and the Rabi oscillation are approximately $-$23$\%$ and $-30\%$, respectively (the negative sign is consistent with the pulsed ODMR contrast discussed below) at room temperature. These outstanding properties are comparable with those of the NV centres in diamond~\cite{Diamond:NV:Single:ODMR:1997, Diamond:NV:Single:CW:ODMR:2011}. The coupling between a single PL6 defect spin and a nearby nuclear spin ($^{29}$Si) is further detected. We also provide a theoretical explanation for the high readout contrast by analysing the defect levels and decay paths in the defects.

To our knowledge, this is the second solid-state defect qubit that exhibits such unique properties in terms of a high readout contrast together with a high photon count rate at room temperature but in a technologically mature material with a wavelength region that is favourable for biological quantum sensing and quantum communication applications. Efficiently generated single divacancy defects in SiC with high-quality room-temperature optical and spin properties would be suitable for nanoscale sensing and helpful for constructing hybrid quantum devices under ambient conditions.\\

\noindent
{\bf EXPERIMENTAL RESULTS}

\begin{figure}[hbt]
\centering
\includegraphics[width=\linewidth]{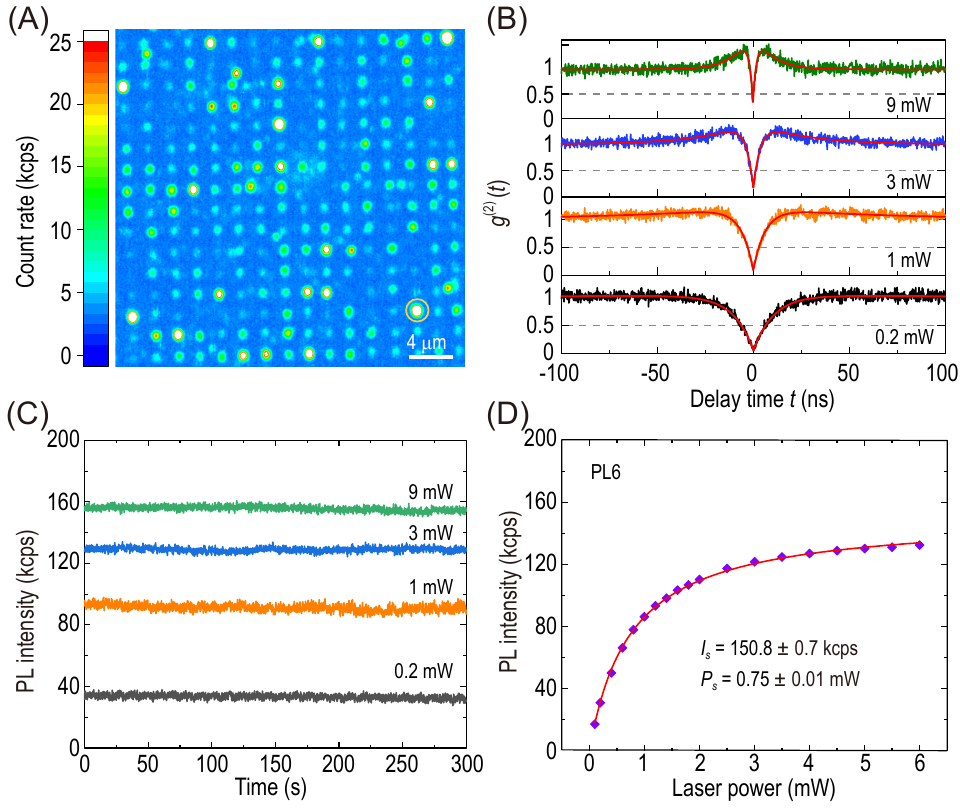}
\caption{{\bf Characterization of the single $\mathrm{V_{Si}V_C}$ defect arrays generated by 30 keV carbon ion implantation.} {\bf (A)} Representative confocal fluorescence image ($32\times32$~$\upmu$m$^2$) of the implanted sample. The white scale bar is 4~$\upmu$m. The bright point in the orange circle represents the single PL6 defect used in {\bf (B)-(D)}. {\bf (B)} Second-order intensity correlation function of $g^{(2)}(t)$ for exciting laser powers of 0.2~mW (black), 1~mW (orange), 3~mW (blue) and 9~mW (green). The red lines are the corresponding fittings. {\bf (C)} Photostability at exciting laser powers of 0.2 mW (black), 1 mW (orange), 3 mW (blue) and 9 mW (green). The sampling time is 0.1 s, and the duration time is 300 s. {\bf (D)} Saturation behaviour. The purple rhombuses are the background-corrected experimental data and the red solid line is the fitting with a function of $I(P)=I_s\cdot P/(P+P_s)$. $P$ and $I(P)$ are the exciting laser power and the corresponding count rate, respectively, with $I_s$ and $P_s$ being the saturated count rate and saturated exciting power, respectively.
}
\label{FigScan}
\end{figure}

\noindent The implanted sample was annealed at 900~$^{\circ}$C for 30 minutes to prepare the single defects (the low-temperature photoluminescence spectra can be found in section 1 in the Supplementary Information (SI)). The detailed process of sample preparation can be found in the Methods section. In the experiment, a 920-nm continuous-wave (CW) laser within the range of optimal excitation wavelengths~\cite{4HSiC:VsiVc:Vsi:ensemble:OCC:01} is used to excite the colour centres. Fig.~\ref{FigScan}A shows a representative confocal fluorescence image within an area of $32\times32$~$\upmu$m$^2$ using home-built confocal microscopy with an oil objective of 1.3~NA. (see Methods for more details). The pumping power is set to 1~mW. In the image, some of the bright points are still shown to be single defects. For example, the defect denoted by the orange circle is a single PL6 defect, which will be investigated in detail later.

\begin{figure*}[bt]
\centering
\includegraphics[width=\linewidth]{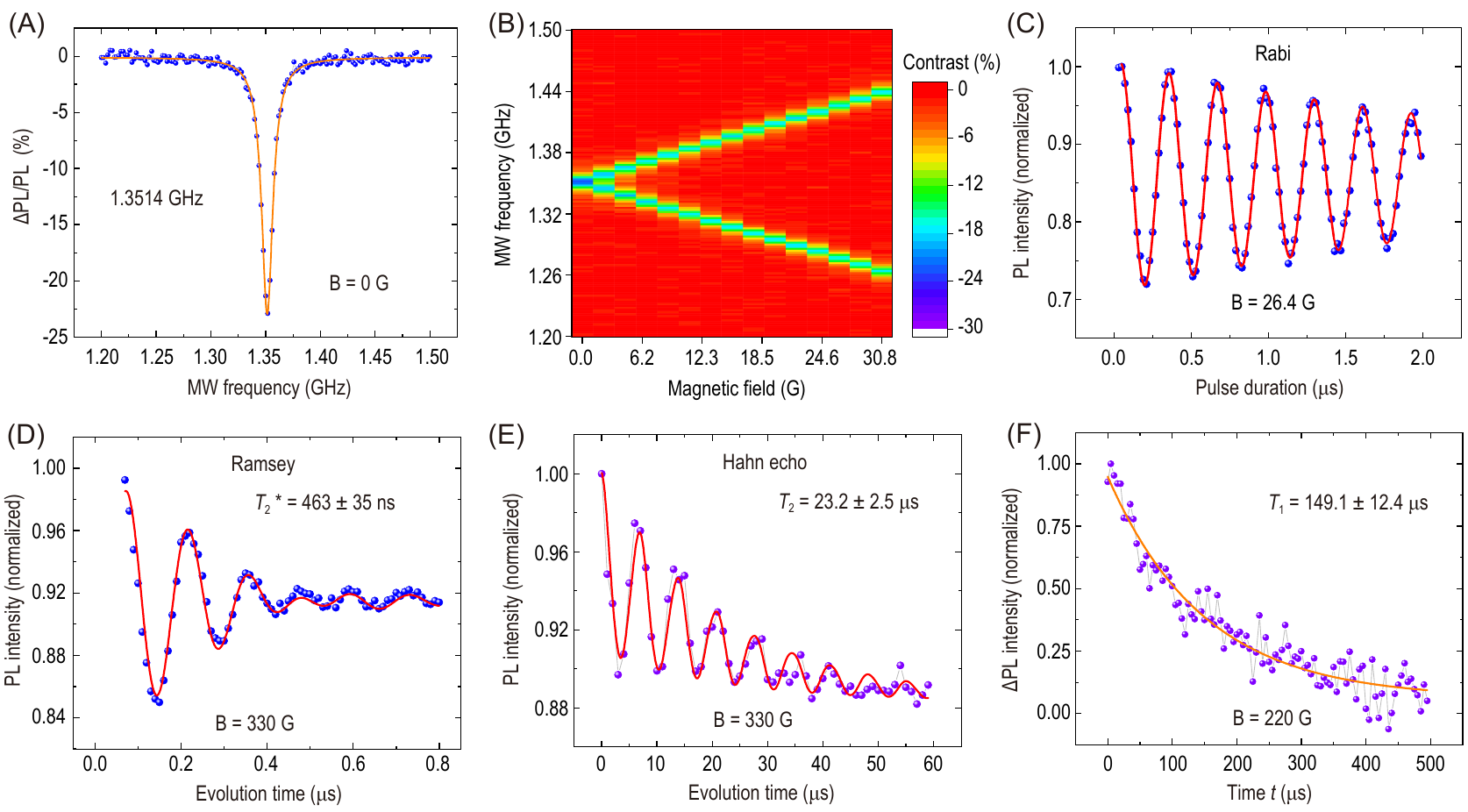}
\caption{{\bf Spin properties of a single PL6 defect at room temperature.} The magnetic field is arranged to be parallel to the crystal $c$-axis. {\bf (A)} CW-ODMR spectra in the zero magnetic field. The blue dots are the experimental raw data, and the orange line is the corresponding Lorentzian fitting centred at 1.3514~GHz. {\bf (B)} The CW-ODMR spectra as a function of the magnetic field intensities. {\bf (C)} Rabi oscillation measured in a magnetic field of 26.4~G. {\bf (D)} Ramsey oscillation measured in a magnetic field of 330~G. From the fitting, the inhomogeneous spin-dephasing time $T_2^*$ is deduced to be $463\pm35$~ns. {\bf (E)} Hahn echo coherence time measured in a magnetic field of 330 G. From the fitting, the homogeneous spin coherence time  $T_2$ is $23.2\pm2.5$~$\upmu$s. {\bf (F)} A representative measurement of $T_1$ with a magnetic field of 220~G. The purple dots are the experimental data, which are fitted by a single-exponential decay function.
}
\label{FigPL6Spin}
\end{figure*}


We characterize the optical properties of the single PL6 defect, denoted by the orange circle in Fig.~\ref{FigScan}A at room temperature. The second-order intensity correlation function is measured at different excitation laser powers (Fig.~\ref{FigScan}B). The obvious photon-bunching phenomenon in the Hanbury-Brown and Twiss (HBT) measurement under the situation of a high exciting laser power implies the existence of a metastable state~\cite{SiC:Diode1}. The background-corrected experimental data are fitted by equation $g^{(2)}(t)=1-(1+a)e^{-|t|/\tau_1}+d e^{-|t|/\tau_2}$, where $a$, $d$, $\tau_1$ and $\tau_2$ are the fitting parameters~\cite{4HSiC:CsiVc:electron:Castelletto, WJF:3CSiC:SPS:Infrared} (the values of $g^{(2)}(t)$ deduced from the raw data are shown in section 2 in the SI for comparison). The values of $g^{(2)}(0)$ at different exciting laser powers are all less than 0.5, indicating a single-photon emitter. We also measured the time traces of the fluorescence intensity of a single PL6 defect with a sampling time of 0.1 s at exciting laser powers of 0.2~mW (black), 1~mW (orange), 3~mW (blue), and 9~mW (green), as shown in Fig.~\ref{FigScan}C. The experimental results indicate that the fluorescence emission of the single PL6 defect at different exciting laser powers is photostable. We further measured its saturation behaviour (Fig.~\ref{FigScan}D). The background-corrected experimental data (purple rhombuses) are fitted with the function $I(P)=I_s\cdot P/(P+P_s)$ (solid red line). The saturated exciting power $P_s$ is 0.75 $\pm$ 0.01~mW, and the saturated PL intensity $I_s$ is 150.8 $\pm$ 0.7~kcps. We also measured the saturated PL intensity of several other randomly selected single PL6 defects. The saturated PL intensity of these single PL6 defects ranges from 138.9~kcps to 172.4~kcps, with an average value of 155.9~kcps (see section 2 in the SI). We observe spin-selective optical lifetimes at 13.4 $\pm$ 0.3~ns and 8.9 $\pm$ 0.1~ns at room temperature (see section 3 in the SI), which implies a sizable ODMR contrast for single PL6 defects at ambient conditions (see the Theoretical Analysis section below).


We then use a common ODMR method (see Methods for more details) to characterize the spin properties of the single PL6 defect at room temperature, which is widely used for NV centres in diamond or divacancies in SiC~\cite{4HSiC:VsiVc:ensemble:Nature:2011, Diamond:Magnetometry:Review}. The CW-ODMR spectrum in the zero magnetic field excited with a 50~$\upmu$W laser is shown in Fig.~\ref{FigPL6Spin}A. Due to the spin polarization-dependent emission, a change is inevitable in the PL readout ($\Delta$PL) with and without the resonant microwave (MW). The oscillation frequency between $m_s=0$ and $m_s=\pm1$ is $1.3514$~GHz~\cite{4HSiC:VsiVc:ensemble:Nature:2011, SiC:VsiVc:ensemble:implantation:C:Awschalom}, and the ODMR contrast is deduced to be $-$23$\%$ (see section 4 in the SI for details on the optimization of ODMR contrast). We further demonstrate the ODMR signals as a function of the magnetic field, which is arranged to be parallel to the crystal $c$-axis (Fig.~\ref{FigPL6Spin}B). The slope of splitting between $m_s=\pm1$ and $m_s=0$ is $\pm2.80$~MHz/G due to the Zeeman effect. The Rabi oscillation of the single spin between $m_s=0$ and $m_s=-1$ states in a magnetic field of 26.4~G is shown in Fig.~\ref{FigPL6Spin}C, where the readout contrast is deduced to be approximately $-30\%$. We also measure the Rabi oscillation contrast of several other single PL6 defects. The Rabi oscillation contrast ranges from $-23.0\%$ to $-31.6\%$ with an average value of $-26.4\%$ (see section 2 in the SI). We then characterized the coherence properties of the single PL6 defect spin at room temperature (see Methods). The Ramsey oscillation is measured in a magnetic field of 330~G, which is shown in Fig.~\ref{FigPL6Spin}D. The experimental data (blue dots) are fitted using a two-cosine exponential decay function (red line), from which the inhomogeneous spin-dephasing time $T_2^*$ is deduced to be 63 $\pm$ 35~ns. The Hahn echo is also measured in a magnetic field of 330~G (Fig.~\ref{FigPL6Spin}E), from which the homogeneous spin coherence time $T_2$ is deduced to be 23.2 $\pm$ 2.5~$\upmu$s. The coherence time can be readily elongated via dynamical decoupling techniques~\cite{Diamond:NV:DD:Hanson:Science2010}. In this work, Carr-Purcell-Meiboom-Gill (CPMG) decoupling sequences~\cite{Diamond:NV:CPMG:2010PRL} are used to prolong the spin coherence time $T_2$ of single PL6 defects. Taking advantage of the CPMG-2 sequences, the coherence time $T_2$ of a selected single PL6 defect spin is extended from $30.2\pm5.5$~$\upmu$s to $41.1\pm3.5$~$\upmu$s. As the number of $\pi$-pulses in the CPMG sequence increases, the coherence time of the electronic spin is extended (see section 5 in the SI for more details). The longitudinal coherence time $T_1$ is further measured to be $149.1\pm12.4$~$\upmu$s in a magnetic field of 220~G, which is shown in Fig.~\ref{FigPL6Spin}F (see section 6 in the SI for the measuring method). The spin coherence time $T_2$ of the single PL6 defect is shorter than that in the as-grown high-purity semi-insulating (HPSI) 4H-SiC~\cite{4HSiC:VsiVc:ensemble:Nature:2011, YFF:4HSiC:VsiVc:ensemble:sensor:HT}, presumably because of the high nitrogen doping level in the used samples ($5\times10^{15}$~cm$^{-3}$) and material damage from the ion implantations~\cite{SiC:VsiVc:ensemble:implantation:C:Awschalom}. $T_2$ can be dramatically improved by using SiC samples with lower nitrogen concentrations and isotopic purification~\cite{4HSiC:Single:Entanglement:NM2020, 4HSiC:Vsi:HOM:20NC}, as well as by optimizing the conditions of implantation and annealing, similar to the strategies usually adopted for NV centres in diamond~\cite{Diamond:NV:LowNitrogen, Diamond:NV:Annealing:CoherenceTime}.

\begin{figure}[hbt]
\centering
\includegraphics[width=\linewidth]{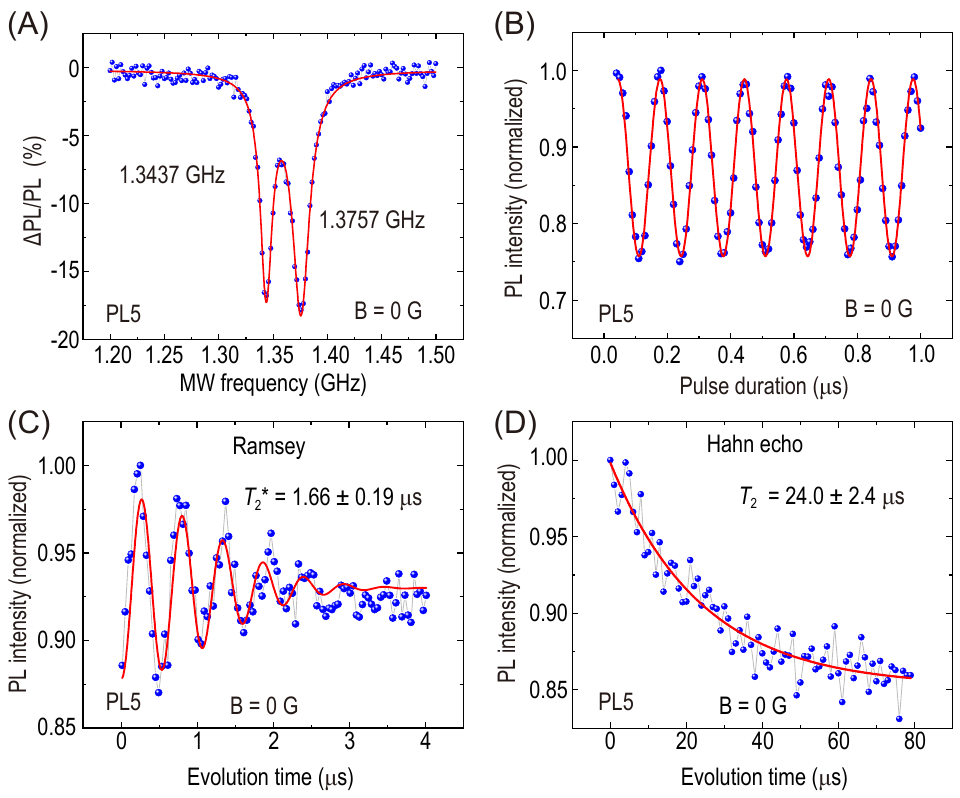}
\caption{{\bf Spin properties of a single PL5 defect spin at room temperature measured in a zero magnetic field.} {\bf (A)} CW-ODMR spectra. The blue dots are the experimental raw data and the red lines represent the corresponding Lorentzian-shaped multipeak fitting. {\bf (B)} Rabi oscillations. The blue dots are the experimental raw data and the red line corresponds to the decaying cosine fittings. \textbf{(C)} Ramsey oscillation. From the fitting, the inhomogeneous spin-dephasing time $T_2^*$ is deduced to be 1.66 $\pm$ 0.19~$\upmu$s. \textbf{(D)} Hahn echo coherence time. From the fitting, the homogeneous spin coherence time $T_2$ is $24.0\pm2.4$~$\upmu$s.
}
\label{FigPL5Spin}
\end{figure}

In this work, we determined the single emitter by measuring the second-order intensity correlation function ($g^{(2)}(t)$) for the isolated bright spots generated by C$^+$ ion implantation and annealing. The values of $g^{(2)}(0)$ deduced from the raw data and the background-corrected results are both far less than $0.5$, indicating a single defect (see SI for more detailed information). In addition, we identified the types of single defects by measuring the ODMR spectra at room temperature or detecting the corresponding fluorescence spectra at a cryogenic temperature of 8~K (see section 1 in the SI). Fig.~\ref{FigPL5Spin} demonstrates the spin properties of a single PL5 defect at room temperature, and Fig.~\ref{FigPL5Spin}A shows the zero-field CW-ODMR spectrum of a single PL5 defect with 50-$\upmu$W laser pumping. The oscillation frequencies are 1.3757~GHz and 1.3437~GHz, respectively~\cite{4HSiC:VsiVc:ensemble:Nature:2011, SiC:VsiVc:ensemble:implantation:C:Awschalom}. It is worth noting that the CW-ODMR contrast of the single PL5 defect spin can approach $-18\%$. We focus on the right branch to investigate coherent manipulation. The Rabi oscillation of the single PL5 defect spin with a zero magnetic field is demonstrated in Fig.~\ref{FigPL5Spin}B, where the readout contrast of the Rabi oscillation is approximately $-25\%$. We also measured the Rabi oscillation of several other single PL5 defects. The Rabi oscillation contrast ranges from $-23.6\%$ to $-28.5\%$, with an average value of $-26\%$ (see section 7 in the SI). The Ramsey and Hahn echo measurements of the single PL5 defect spin are demonstrated in Fig.~\ref{FigPL5Spin}C and Fig.~\ref{FigPL5Spin}D, respectively. From the fitting, the inhomogeneous spin-dephasing time $T_2^*$ and the Hahn echo coherence time $T_2$ of the single PL5 defect without a magnetic field at room temperature are deduced to be $1.66\pm0.19$~$\upmu$s and $24.0\pm2.4$~$\upmu$s, respectively. More optical and spin properties of the single PL5 defect can be found in section 7 in the SI.

We also investigated the spin properties of single PL1 and PL7 defects at room temperature (see section 8 in the SI). The contrasts of Rabi oscillation of single PL1 and PL7 are approximately $-6.6\%$ and $-10\%$, respectively, which are both lower than those of PL5 and PL6. The properties of PL1-7 are summarized in Table S1 in the SI. We further measured the generation ratio of single PL5, PL6, and PL7 defects in a $10\times10$ array of implanted sites, which are obtained to be 7\%, 1\%, and 6\%, respectively (see section 9 in the SI).\\

\begin{figure}[hbt]
\centering
\includegraphics[width=\linewidth]{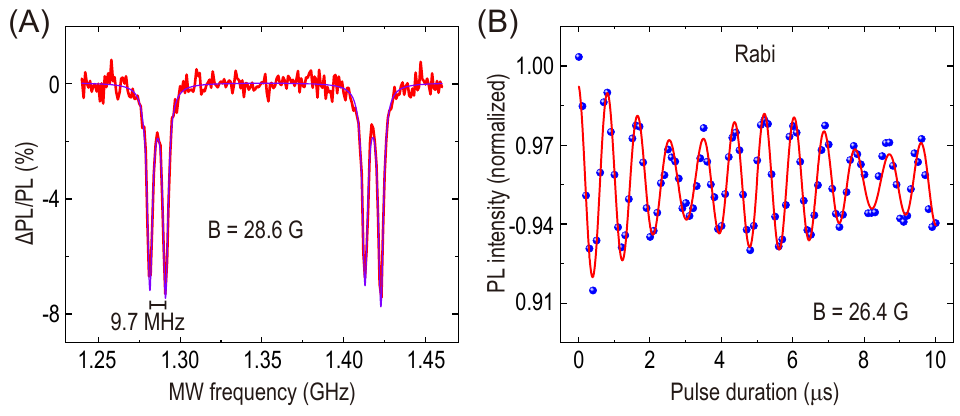}
\caption{\textbf{A single PL6 electron spin coupled to a nearby $^{29}$Si nuclear spin at room temperature.} \textbf{(A)} ODMR spectra in a magnetic field of 28.6~G. \textbf{(B)} Rabi oscillation of the defect spin hyperfine coupling with a nuclear spin.
}
\label{FigNuclearSpin}
\end{figure}

For the SiC sample with natural abundance, there are 4.7$\%$ $^{29}$Si with nuclear spin $I_{\mathrm{Si}}=1/2$ and 1.1$\%$ $^{13}$C with nuclear spin $I_{\mathrm{C}}=1/2$. In the implanted sample, it is easy to find a single defect spin strongly coupled with a nearby nuclear spin, even at room temperature. The ODMR spectra of a single PL6 defect spin coupled with a nearby $^{29}$Si nuclear spin (Si$_{\mathrm{\uppercase\expandafter{\romannumeral2}}\mathrm{b}}$ lattice site)~\cite{4HSiC:VsiVc:ensemble:entanglement, 4HSiC:6HSiC:NuclearSpin:Sensor:Magnetic} are measured in a $c$-axis magnetic field of 28.6~G, which is shown in Fig.~\ref{FigNuclearSpin}A. The splitting of two sets of dips results from the Zeeman effect with the 9.7~MHz splitting derived from the hyperfine interaction remaining consistent. Fig.~\ref{FigNuclearSpin}B shows the Rabi oscillation between the $|0_e\downarrow_n\rangle$ and $|-1_e\downarrow_n\rangle$ states ($|0_e\rangle$ and $|-1_e\rangle$ represent the electron spin states of $m_s=0$ and $m_s=-1$, respectively, and $|\downarrow_n\rangle$ represents the nuclear state of $m_I=-1/2$). The blue dots denote the experimental data and are fitted by a two-cosine exponential decay function (red solid line).\\


\noindent
{\bf THEORETICAL ANALYSIS}

The room temperature readout contrast of the PL5 and PL6 centres in 4H-SiC is strikingly high, which requires a theoretical interpretation. The ODMR contrast of divacancy defects in 4H-SiC can be analysed based on the theory of the ODMR contrast of the NV centre in diamond because they are isovalent centres~\cite{SiC:VsiVc:Theory1:Gali:2011,4HSiCand3CSiC:VsiVc:single:02}. The so-called $c$-axis divacancy defects, in which the neighbouring carbon and silicon vacancies are situated along the $c$-axis of the crystal, possess $C_{3v}$ symmetry similar to the diamond NV centre. The basal divacancy defects exhibit $C_{1h}$ symmetry in 4H-SiC, but it has been recently shown~\cite{Science_Advances_5_eaay0527} that these configurations can be considered as $C_{3v}$ defects with a spin quantization axis pointing along the connecting line of the vacancies with a perturbation of strain induced by the crystal field. By considering the strain as a relatively small perturbation, one can focus on the $C_{3v}$ symmetry solution as obtained for the diamond NV centre.

\begin{figure}[hbt]
\centering
\includegraphics[width=0.9\linewidth]{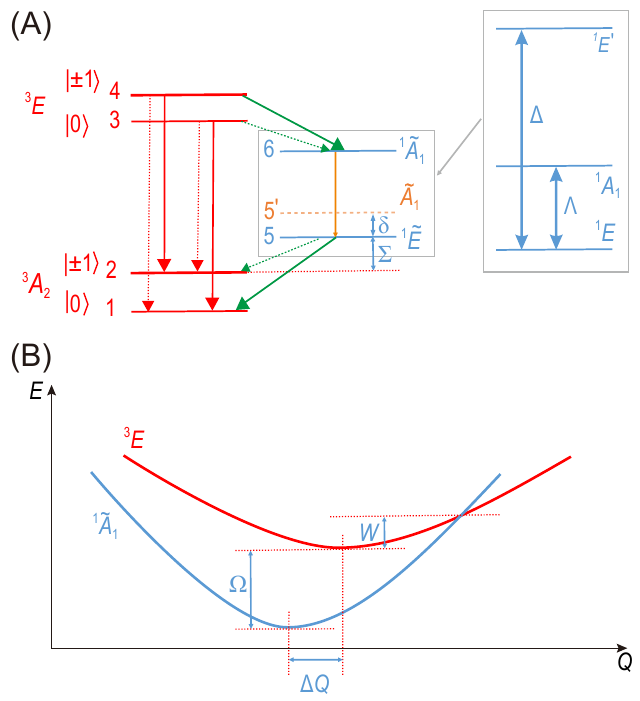}
\caption{\textbf{Theoretical model.} {\bf (A)} Defect levels and decay paths in divacancy defects. This simplified diagram is valid at room temperature in the ${}^3E$ excited state and at zero strain and magnetic fields. The radiative decays are depicted by the red arrows, whereas the non-radiative processes are shown by the green arrows. Radiative and non-radiative decays compete between the ${}^1\widetilde{A}_1$ and ${}^1\widetilde{E}$ singlet states (see the box in the middle of the figure), as depicted by an orange arrow. The very weak transitions are depicted by thin dotted arrows. The tilde label represents vibronic states, in which strong electron-phonon coupling mixes the three singlet states, as shown in the box on the right side of the figure. The $\widetilde{A}_1$ state (orange dashed level) is the first excited vibronic state over the ${}^1\widetilde{E}$ ground state, which plays a role in the temperature dependence of intersystem crossing towards the ground state manifold. The defect states are enumerated. The corresponding energy gaps are labelled by Greek letters. The energy gaps between the $ms=\pm1$ and $ms=0$ spin levels are magnified by six orders of magnitude for the sake of clarity. {\bf (B)} Levels crossing between the bright and dark excited states. Schematic energy ($E$) -- configuration coordinate ($Q$) diagram for the divacancy defects. At cryogenic temperatures, the energy gap is $\Omega$ between the bright ${}^3E$ and dark ${}^1\widetilde{A}_1$ states. The barrier energy for the ${}^3E$ state is $W$ in order to reach the crossing point between the two levels.
}
\label{FigLevel}
\end{figure}

The analysis of the ODMR contrasts is based on the known levels and states of the defect (see Fig.~\ref{FigLevel}A and Refs.~\cite{Nanophotonics_8_1907, npjQuantMat_3_31} and references therein) that are labelled according to the $C_{3v}$ point group. The levels are enumerated for the sake of simplicity. The ODMR contrast depends on the relation between the intersystem crossing (ISC) rates (green arrows) and the direct recombination rates (red arrows) as the electron decays from the excited state manifold (states 4 and 3) to the ground state manifold (states 1 and 2) via the metastable states (states 6 and 5) and directly, respectively. The strength of the transitions is governed by selection rules and electron-phonon coupling where the latter results in vibronic singlet states labelled by a tilde in Fig.~\ref{FigLevel}A (see section 10 in the SI for details).

The observed ODMR readout contrast depends on a number of factors. To consider the trends, we simplify this complex problem to an expression with parameters that are intrinsic to the defects. In this case, the pulsed off-resonant ODMR readout contrast $C$ can be expressed as
\begin{equation}
\label{eq:odmr}
C = \frac{\tau_{\pm1}-\tau_0}{\tau_0} =
\frac{r_0 - r_{\pm1}}{r_{\pm1}} \text{,}
\end{equation}
where $r_0=r_{D}+r_{36}$ and $r_{\pm1}=r_{D}+r_{46}$ are the corresponding rates with $r_D$ direct recombination rate, $r_D=r_{42}=r_{31}$, and the respective $\tau_0$ and $\tau_{\pm1}$ are the optical lifetimes (inverse of the rates). The rates $r_{36}$ and $r_{46}$ are the corresponding ISC rates, where $r_{36}$ is extremely weak and can be ignored~(see Ref.~\cite{Nanophotonics_8_1907} and references therein). As a consequence, the sign of the ODMR contrast will be negative, as $r_{\pm1} > r_{0}$ applies in this condition. Eq.~\eqref{eq:odmr} rests upon four basic assumptions: (i) perfect optical spin-polarization of state 1 ($m_s = 0$ ground state) upon long illumination before the readout protocol starts, i.e., perfect initialisation; (ii) photo-excitation of state 1 will preserve the $m_s = 0$ state, i.e., $r_{36}$ is negligible so the reference fluorescence intensity is the fluorescence of $m_s = 0$ state (state 3); (iii) perfect spin-flip upon applying a microwave $\pi$-pulse in state 1 to rotate it to state 2, so the change in fluorescence intensity is associated with emission from $m_s = \pm1$ state (state 4); (iv) the change of fluorescence intensity can be perfectly measured, i.e., it is associated with the optical lifetime of $m_s = \pm1$ state (state 4). We discuss in section 10 in SI how conditions (i)-(iii) are fulfilled from the theoretical point of view. In practice, the measurements are not perfect. In particular, they often fail to satisfy condition (iv). As a consequence, the observed ODMR readout contrasts weaker than the theoretical limit that is an upper bound for the absolute value of the pulsed ODMR readout contrast. We also emphasize that the deviation from the theoretical limit may strongly vary depending on the applied parameters and technicalities in the actual measurements, even for the same type of divacancy defect. This makes the direct comparison between studies with different readout parameters or experimental setups ambiguous. Thus our theory is used to interpret the trends in the observed ODMR contrasts but is not intended to fully comply with the experimental data.

As an example, we apply this theory for the room temperature data for a single PL6 and PL1 centre are recorded in Fig.~S5 in the SI. By taking the average lifetime data, one obtains $-33.6$\% and $-10.0$\% ODMR readout contrasts, respectively, that are also close to the observed room temperature ODMR readout contrasts (average values are $-26.4$\% and $-6.6$\%, respectively). We note that the observed lifetimes show non-negligible uncertainties which affect the results on the derived ODMR readout contrasts. Nevertheless, it may be concluded that the theoretical upper bound limit is indeed manifested. We note that observations on single PL6 centres approach the theoretical limit (31.6\%) for one PL6 defect in our study (see Fig.~S4B). These results also imply that Eq.~\eqref{eq:odmr} can be applied to understand the trends in the ODMR contrasts either in terms of various divacancy configurations or temperature dependence of a given divacancy configuration in 4H-SiC, and the dissimilarities between the diamond NV centre and SiC divacancies are quantitative and not qualitative.

The ODMR contrast is primarily governed by the rate $r_{46}$. The rate $r_{46}\propto \lambda_\perp^2 F_{A}(\Omega)$ depends on the strength of the spin-orbit coupling (perpendicular component, $\lambda_\perp$) and the spectral phonon overlap function $F_A$ with the $A_1$ phonons that connect the geometries of $\left|^1\widetilde{A}_1\right\rangle$ and $\left|^3E\right\rangle$ that are originally separated by $\Delta Q$ (see Fig.~\ref{FigLevel}B). The spin-orbit coupling parameters are on the same order of magnitude for the diamond NV centre and the divacancy defects in SiC (see Refs.~\onlinecite{Nanophotonics_8_1907, 4HSiCand3CSiC:VsiVc:single:02} and the references therein); nevertheless, $\lambda_\perp$ values may differ somewhat for the diamond NV centre and the divacancy configurations, which can contribute to quantitative differences in the final values of $r_{46}$. On the other hand, the $\lambda_\perp$ values should be very similar for each divacancy configuration because of the common chemical composition. It is likely that the quantitative differences between the $r_{46}$ rates and the corresponding ODMR readout contrasts of the defects dominantly come from the strongly varying $F_A$ values.

The $F_A$ values are sensitive to the low-temperature energy gap between the excited state triplet and the nearby singlet ($\Omega$ in Fig.~\ref{FigLevel}B; see discussion in section 10 in the SI). $F_A$ rapidly increases by closing the gap $\Omega$, which ultimately results in a larger ODMR readout contrast. Since the electronic states are confined for PL5 and PL6 in the stacking faults (see section 10 in the SI), $\Omega$ is expected to be smaller for the PL5 and PL6 defects than for the PL1-4 defects which explains the trends in the low-temperature ODMR contrasts of divacancy configurations in 4H-SiC. While yielding the room temperature ODMR contrast requires understanding the temperature dependence of the ODMR contrast, which depends on the temperature dependence of $\tau_{0}$ and $\tau_{\pm1}$. In a seminal work, it has been found for a single diamond NV centre~\cite{PRX_2_031001} that $\tau_{\pm1}$ is almost constant over a wide range of temperatures ($300\dots680$~K); however, $\tau_{0}$ radically decreased for temperatures above $550$~K. This can be interpreted such that $r_{36}$, the nonradiative decay from the triplet $\left|0\right\rangle$, is significantly enhanced at elevated temperatures. They provided a phenomenological model to explain this phenomenon, the Mott-Seitz formula, which was developed for multi-phonon non-radiative processes. Fig.~\ref{FigLevel}B shows that the bright and dark levels can cross by the dynamic motion of ions. The energy barrier $W$ jumps into this crossing point from the lowest energy of the bright state, which can be reached at elevated temperatures $T$ with the thermal energy $k_\text{B}T$, where $k_\text{B}$ is the Boltzmann-constant. In this case, the lifetime of state 3, {$\tau_0(T)$}, can be expressed as

\begin{equation}
    \label{eq:MottSeitz}
    \tau_0(T) = \frac{\tau_0(T\approx0 \text{K})}{1 + s \times \exp{\left(-\frac{W}{k_\text{B}T}\right)}} \text{,}
\end{equation}
where $s$ is a dimensionless quantity, and is interpreted as the fraction of the non-radiative and radiative rates at the crossing point.
The energy $W\approx0.5$~eV is relatively large for the diamond NV centre~\cite{PRX_2_031001, npjQuantMat_3_31}; therefore, the strong temperature dependence on the ODMR readout contrast is only visible from $T\approx550$~K. However, the energy gaps for divacancy defects in 4H-SiC are much smaller. For instance, $W\approx0.1$~eV is calculated for the PL1 defect~\cite{npjQuantMat_3_31}. This result implies that the threshold temperature is much lower for divacancy defects than for the diamond NV centre.

No readable quantitative data is available in the literature about the temperature dependence of the pulsed ODMR readout contrast of divacancy defects in SiC, and our paper focuses on the room temperature properties. On the other hand, low-temperature off-resonant ODMR contrast data is available on ensemble defects by measuring only the zero-phonon lines~\cite{4HSiC:VsiVc:ensemble:sensor:electrical:strain}. Because of the very different conditions of measurements, the raw CW-ODMR contrast data (e.g., $\approx-10$\% for the PL1 defects) in Ref.~\onlinecite{4HSiC:VsiVc:ensemble:sensor:electrical:strain} should be scaled up by a factor ($\approx1.444$) that we estimated from our data by comparing the CW-ODMR contrast and the pulsed ODMR contrast at room temperature (see details in section 10 in the SI). The estimated pulsed ODMR contrast data with the experimental conditions in our setup and protocol corresponds to $-14.5$\%\R{,} which reduces to $\approx-6.6$\% at room temperature (see section 8 in the SI in this work). Thus, the ODMR contrast of divacancy defect significantly weakens at elevated temperatures. By using the same procedure with using data from Ref.~\onlinecite{4HSiC:VsiVc:ensemble:sensor:electrical:strain}, the deduced pulsed ODMR contrasts of PL5 and PL6 defects associated with our experimental setup and protocol are $\approx-34.7$\% and $\approx-31.8$\%, respectively. The reduction in the contrasts is not insignificant up to room temperature, but the absolute values of the contrasts still remain relatively high for single PL5 and PL6 defects ($\approx-26$\% and $\approx-26.4$\%, respectively, average values in this work), and may approach zero above $650$~K~\cite{YFF:4HSiC:VsiVc:ensemble:sensor:HT}. Indeed, our theory can excellently account for the temperature-dependent ODMR readout contrast of PL5 defects observed at a wide temperature range, resulting in $W=0.076\pm0.003$~eV from the fit to data extracted from experiments (see section 10 in the SI). The value of $W$ for PL5 defect is indeed smaller than the calculated $W=0.1$~eV for the PL1 defect, which is consistent with our quantum confinement theory on the defect levels of divacancy defects in the stacking fault (see section 10 in the SI).\\

\noindent
{\bf CONCLUSION}

In conclusion, we presented a scalable method for the creation of single divacancy spin defects in 4H-SiC using carbon ion implantation and combining electron beam lithography and post-annealing techniques. We characterized the spin properties and demonstrated the coherent manipulation of individual spin defects, including PL1, PL5, PL6 and PL7 defects, at room temperature. Surprisingly, single PL6 spin defects have some outstanding properties compared with several previously reported spin defects in SiC~\cite{4HSiC:VsiVc:ensemble:Nature:2011, 4HSiC:VsiVc:single:01, 4HSiCand3CSiC:VsiVc:single:02,4HSiC:Vsi:single:SIL:RT:ODMR:PL:2015, 4HSiC:Vsi:Qudits:RT}. The saturated count rate of a single PL6 centre is up to 150~kcps, and its CW-ODMR and Rabi oscillation contrasts can reach as high as $-23\%$ and $-30\%$, respectively, which are comparable with those of single NV colour centres in diamond. By analysing the defect levels and decay paths, we provide a theoretical model to explain the observed high ODMR contrast, in which the experimental results agree well with the theoretical predictions.

The divacancy qubits reported in this work have near-infrared excitation and emission in a wavelength region that is the most transparent to living cells; this is in stark contrast to the NV centres in diamond which require green illumination for efficient photo-excitation, causing high auto-fluorescence of living cells. Besides, the longer laser wavelength required for ODMR measurements of colour centres in SiC is advantageous in biological applications, compared with the NV centres in diamond, regarding the photo-toxicity. This makes divacancy colour centres highly prospective for biological and human diagnostic applications and therapy as similar divacancy defects have been engineered into water soluble SiC nanocrystals~\cite{SiC:Nanocrystals:Qubits}. Integrating spin defects with a high readout contrast and a high photon count rate into high-performance SiC electron devices and recently developed integrated optical chips based on SiC~\cite{SiC:IntegaratedOptics:2019:NPhoton} may also provide considerable opportunities for the next generation of hybrid quantum devices.

\vbox{}
\noindent
\noindent
{\bf Methods}

\noindent
\textbf{Sample preparation.}
In our work, a 12.5-$\upmu$m-thick epitaxial layer of single-crystal 4H-SiC with a nitrogen doping density of $5\times10^{15}$ cm$^{-3}$ grown on a 4$^\circ$ off-axis 4H-SiC substrate was used~\cite{LQ:4HSiC:Vsi:single:array,WJF:4HSiC:NV:PRL2020}. A layer of positive electron beam photoresist PMMA A4 with a thickness of approximately 200~nm was spin-coated onto the surface of the SiC sample. Through electron beam lithography (EBL, JEOL, JBX 6300FS), an array of apertures with a pitch of 2~$\upmu$m and a diameter of $50\pm10$~nm was fabricated on the surface of the sample as a mask. Then, the sample was implanted with 30-keV C$^+$ ions at a dose of $1.02\times10^{12}$~cm$^{-2}$. There were approximately $20$ implanted carbon ions per aperture in the sample. The mask was then removed using an ultrasonic bath of acetone solution. The sample was annealed in a tube furnace at 900~$^{\circ}$C for $30$ minutes in a vacuum environment of approximately $1\times10^{-4}$ Pa. Finally, the sample was cleaned in a 3:1 mixture of concentrated sulfuric acid and hydrogen peroxide and heated to 95~$^{\circ}$C for 5 hours, which dramatically reduced the background fluorescence. Single spin defects could then be optically addressed.

\textbf{Optical measurements.}  A home-built scanning confocal microscope with an infrared oil objective with an NA of 1.3 (Nikon, CFI Plan Fluor 100X Oil) was used in our experiments. In all of the optical measurements, a 920-nm CW laser, filtered by a shortpass filter (Thorlabs, FESH950), was used to excite those colour centres. A dichroic beamsplitter (Semrock, Di02-R980-25$\times$36) was then used to separate the laser and fluorescence signals. For various measurements at room temperature, the SiC samples were mounted on a closed cycle three-axis piezoelectric stage (PI, E-727.3SD). The fluorescence signals filtered by a 1000-nm longpass filter (Thorlabs, FELH1000) were coupled to a single mode fibre and then guided to a superconducting nanowire single photon detector (SNSPD, Scontel $\&$ Photon Technology) with an approximately 80$\%$ quantum efficiency. The number of photons is recorded by a counter (NI, USB-6341). For the HBT measurements, the fluorescence signals were divided by a fibre beam splitter and detected using a two-channel SNSPD. The coincidence correlation with variable delay time $t$ was measured using a time-to-digital converter (IDQ, ID800-TDC).

\textbf{Spin coherent manipulation.} The same home-built scanning confocal microscope was used to polarize and readout the optical signals depending on the spin states of the isolated defects. For the ODMR, Rabi, Ramsey and spin echo measurements, the microwave sequences were generated using a synthesized signal generator (Mini-Circuits, SSG-6000 RC) and then gated by a switch (Mini-Circuits, ZASWA-2-50DR+). After amplification by an amplifier (Mini-Circuits, ZHL-25W-272+), the microwave signals were fed to a 50-$\upmu$m-wide copper wire above the surface of 4H-SiC sample. The exciting 920-nm CW laser was modulated using an acousto-optic modulator. The timing sequence of the electrical signals for manipulating and synchronizing the laser, microwave and counter was generated using a pulse generator (SpinCore, PBESR-PRO500).



\noindent
{\bf  Acknowledgments}\\
\noindent
We thank Gang-Qin Liu from Institute of Physics, Chinese Academy of Sciences for his helpful discussion. This work was supported by the National Key Research and Development Program of China (Grant No.\ 2016YFA0302700), the National Natural Science Foundation of China (Grants No.\ U19A2075, 61725504, 61905233, 11774335, 11821404 and 11975221), the Key Research Program of Frontier Sciences, CAS (No. QYZDY-SSW-SLH003), Science Foundation of the CAS (ZDRW-XH-2019-1), Anhui Initiative in Quantum Information Technologies (AHY060300 and AHY020100), the Fundamental Research Funds for the Central Universities (Grants No.\ WK2030380017 and WK2470000026), the National Postdoctoral Program for Innovative Talents (Grant No.\ BX20200326). A.~G. acknowledges the support from the National Research, Development and Innovation Office of Hungary (NKFIH) for Quantum Technology Program (Grant No.\ 2017-1.2.1-NKP-2017-00001), National Excellence Program (Grant No.\ KKP129866), the EU QuantERA Nanospin project (NKFIH Grant No.\ NN127902) as well as the Quantum Information National Laboratory sponsored by the Ministry of Innovation and Technology of Hungary. This work was partially carried out at the USTC centre for Micro and Nanoscale Research and Fabrication.
\\

\noindent
{\bf Author Contributions}\\
\noindent
Q. L., J.-F. W. and J.-S. X. designed the experiments. Q. L. and H. L. prepared the PMMA mask with the support of G.-P. G. L.-P. G. and X. Z. performed the ion implantation experiments. Q. L. carried out the experiments assisted by J.-F. W., F.-F. Y., J.-Y. Z., H.-F. W., Z.-H. L., Z.-Q, W., K. S., J.-S. T. and J.-S. X. A. G. provided the theoretical support. J.-S. X., C.-F. L. and G.-C. G. supervised the project. Q. L., J.-S. X. and A. G. wrote the paper with input from other authors. All authors discussed the experimental procedures and results.

\noindent
{\bf Competing financial interests}\\
\noindent
The authors declare no competing financial interests.

\end{document}